\pdfoutput=1

\documentclass[11pt]{article}
\usepackage{xcolor}
\usepackage{graphicx}
\usepackage{subfigure}
\usepackage{stfloats}
\usepackage{color}
\usepackage{pifont}
\usepackage{balance}

\usepackage[]{acl}
\usepackage{times}
\usepackage{latexsym}
\usepackage{amsmath}
\usepackage{amssymb}

\usepackage[T1]{fontenc}

\usepackage[utf8]{inputenc}
\usepackage{xcolor}

\usepackage{microtype}
\usepackage{xspace}
\usepackage{inconsolata}

\usepackage{multirow}
\usepackage{graphicx}

\newcommand{\benchmark}{\textsc{DebugEval}\xspace}

\title{COAST: Enhancing the Code Debugging Ability of LLMs through Communicative Agent Based Data Synthesis}

\author{
 \textbf{Weiqing Yang\textsuperscript{1}},
 \textbf{Hanbin Wang\textsuperscript{2}},
 \textbf{Zhenghao Liu\textsuperscript{1}\thanks{indicates corresponding author.}},
 \textbf{Xinze Li\textsuperscript{1}},
 \textbf{Yukun Yan\textsuperscript{3}},
\\
 \textbf{Shuo Wang\textsuperscript{3}},
 \textbf{Yu Gu\textsuperscript{1}},
 \textbf{Minghe Yu\textsuperscript{4}},
 \textbf{Zhiyuan Liu\textsuperscript{3}} \textbf{and}
 \textbf{Ge Yu\textsuperscript{1}}
\\
 \textsuperscript{1}Department of Computer Science and Technology, Northeastern University, China
\\
 \textsuperscript{2}School of Software and Microelectronics, Peking University, China
\\
 \textsuperscript{3}Department of Computer Science and Technology, Institute for AI, Tsinghua University, China
\\
 \textsuperscript{4}Software College, Northeastern University, China
}

\begin{document}
\maketitle

\begin{abstract}

Code debugging is a vital stage of software development, essential for ensuring the reliability and performance of Large Language Models (LLMs) in the code generation task. Human debugging typically follows a multi-stage process, which includes Bug Localization, Bug Identification, Code Repair, and Code Recognition. However, existing code debugging benchmarks predominantly focus on the Code Repair stage, which offers only a limited perspective on evaluating the debugging capabilities of LLMs. In this paper, we introduce \benchmark, a comprehensive benchmark for evaluating the debugging abilities of LLMs by emulating the multi-stage human debugging process. Through evaluating on \benchmark, we observe that 7B-scale models consistently underperform compared to their larger counterparts, highlighting their limitations in comprehending code semantics. In this case, we propose the \textbf{CO}mmunicative \textbf{A}gent-based data \textbf{S}yn\textbf{T}hesis (COAST) framework, which employs a multi-agent system to generate high-quality training data for supervised fine-tuning (SFT). Experimental results demonstrate that COAST-generated data outperform human-curated and GPT-4-generated data, enabling 7B-scale LLMs to achieve debugging performance comparable to GPT-3.5. All data and codes are available at \url{https://github.com/NEUIR/COAST}.
\end{abstract}

\section{Introduction}
In software development, code debugging is a crucial stage for ensuring the functionality and reliability of applications~\cite{hailpern2002software,kirschner2020debugging}. Traditional code debugging methods often rely on heuristics~\cite{Le_Goues_Nguyen_Forrest_Weimer_2012,Wen_Chen_Wu_Hao_Cheung_2018} and predefined patterns~\cite{Hua_Zhang_Wang_Khurshid_2018,Liu_Koyuncu_Kim_Bissyandé_2019}. However, these methods are reaching their limits, as software systems become increasingly complex. 

\begin{figure}[t] \centering
    \includegraphics[width=0.49\textwidth]{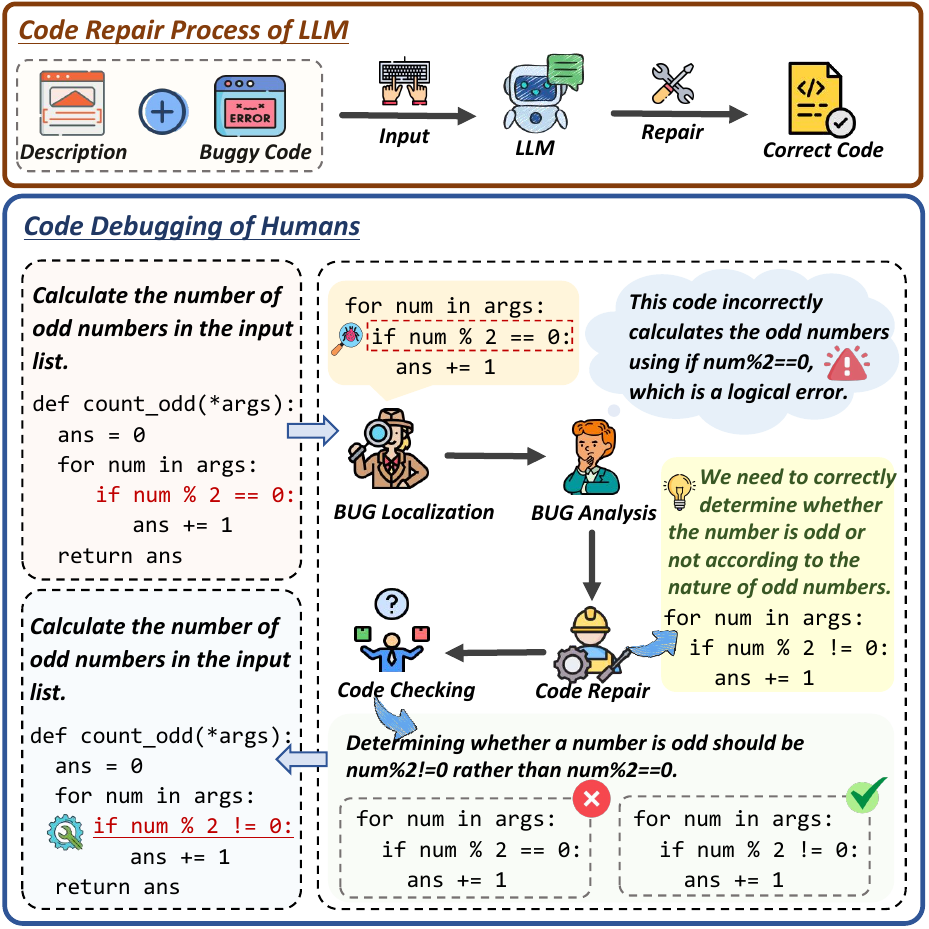}
    \caption{Illustration of the Human Code Debugging Process. Existing studies~\cite{olausson2023self,chen2023teaching} typically focus on directly repairing code generated by Large Language Models (LLMs). In contrast to this approach, humans often engage in a multi-stage process to resolve buggy codes.} 
    \label{fig: debug}
\end{figure}
Large Language Models (LLMs)~\cite{chatgpt,touvron2023llama} have opened new avenues for automated code debugging, enabling more flexible and comprehensive methods to identify and resolve code errors~\cite{chen2023teaching}. As illustrated in Figure~\ref{fig: debug}, the debugging process of humans typically involves multiple essential stages: locating buggy code segments, analyzing root causes of bugs, and repairing identified issues~\cite{chen2406coder}. In addition, a comparison is made between the code before and after repair to check whether the issue is actually solved. Each of these stages is critical for successfully fixing bugs. However, existing debugging benchmarks~\cite{tian2024debugbench,khan2023xcodeeval,huq2022review4repair} primarily focus on assessing LLMs' capability to fix bugs, overlooking their performance in the distinct stages in the human debugging process.

To facilitate more comprehensive evaluations, this paper introduces \benchmark, a benchmark specifically designed to assess the code debugging capabilities of LLMs. \benchmark introduces four tasks--BUG Localization, BUG Identification, Code Repair, and Code Recognition--that closely mirror the debugging process of humans in the real world. Each task spans multiple programming languages, including Python, C++, and Java. To better simulate real-world software development, the buggy codes in \benchmark are from human and GPT-4. We evaluate the debugging performance of various LLMs on \benchmark and observe that LLMs with 7 billion parameters exhibit significantly weaker debugging capabilities compared to LLMs with 70 billion parameters or more. Thus, improving the debugging proficiency of LLMs themselves remains a critical challenge for building more comprehensive code intelligence.

While Supervised Fine-Tuning (SFT) has been widely adopted to enhance LLMs' performance for specialized tasks using human-labeled or LLM-generated data~\cite{roziere2023code,yue2023disc,zhang2023self}, its effectiveness is often limited by the quality of the labeled data. To improve the quality of synthesized data, we propose the \textbf{CO}mmunicative \textbf{A}gent-based data \textbf{S}yn\textbf{T}hesis (COAST) framework. COAST builds three agents for collaboration: \texttt{Code Quizzer}, \texttt{Code Learner}, and \texttt{Code Teacher}. These agents work together to generate code debugging data for finetuning the \texttt{Code Learner}. Specifically, the \texttt{Code Quizzer} first creates a diverse range of code debugging problems. The \texttt{Code Learner} then attempts to answer these questions, serving as a critic to assess their educational value. Problems that the \texttt{Code Learner} answers incorrectly are flagged and curated as SFT data. Finally, the \texttt{Code Teacher} enriches these problems by providing detailed explanations and guidance. The synthesized data are collected and used to finetune the \texttt{Code Learner}, enabling the development of our NeuDebugger model.

Our experiments on \benchmark demonstrate that the COAST framework significantly enhances the debugging capabilities of 7B-scale LLMs. Notably, further analysis reveals that collecting SFT data from human trials and LLM solely does not improve the debugging performance of LLMs~\cite{gudibande2024false}. In the COAST framework, the \texttt{Code Teacher} effectively guides the \texttt{Code Learner} through three tasks--BUG Localization, BUG Identification, and Code Recognition--by employing Chain-of-Thought (CoT) reasoning~\cite{wei2023chainofthoughtpromptingelicitsreasoning}. However, CoT reasoning negatively affects performance in the Code Repair task, which introduces noise and disrupts the underlying code structure. Additionally, the synthesized data significantly boost the performance of code-focused LLMs, such as DeepSeek-Coder-6.7B-Ins, more effectively than general-purpose LLMs like Llama3-8B-Ins. These findings provide valuable insights for future research that aims to improve LLMs' debugging capabilities.

 \section{Related Work}

Debugging is a critical stage to ensure the quality of code generated by Large Language Models (LLMs)~\cite{olausson2023self,chen2023teaching}. Early approaches primarily relied on feature-based methods, such as templates~\cite{Hua_Zhang_Wang_Khurshid_2018,Liu_Koyuncu_Kim_Bissyandé_2019}, heuristic rules~\cite{Le_Goues_Nguyen_Forrest_Weimer_2012,Wen_Chen_Wu_Hao_Cheung_2018}, or constraints~\cite{Mechtaev_Yi_Roychoudhury_2016,DeMarco_Xuan_Le_Berre_Monperrus_2014}, to repair buggy code. However, these methods often fall short in addressing a wide range of bug types or tackling more complex programming challenges.

With the advancement of the pretraining technique, researchers have begun to explore the application of Pretrained Language Models (PLMs) to automated code debugging. For example, \citet{xia2022practical} utilize the code-oriented pretrained model CodeX~\cite{chen2021evaluatinglargelanguagemodels} to investigate the debugging potential of PLMs. Their findings reveal that CodeX excels at repairing code, particularly for Python and Java. Similarly, \citet{kolak2022patch} employ GPT-2~\cite{radford2019language} and CodeX to evaluate their ability to generate accurate patches for buggy code lines based on given prefixes. 

\begin{figure*}[t] \centering
    \includegraphics[width=\textwidth]{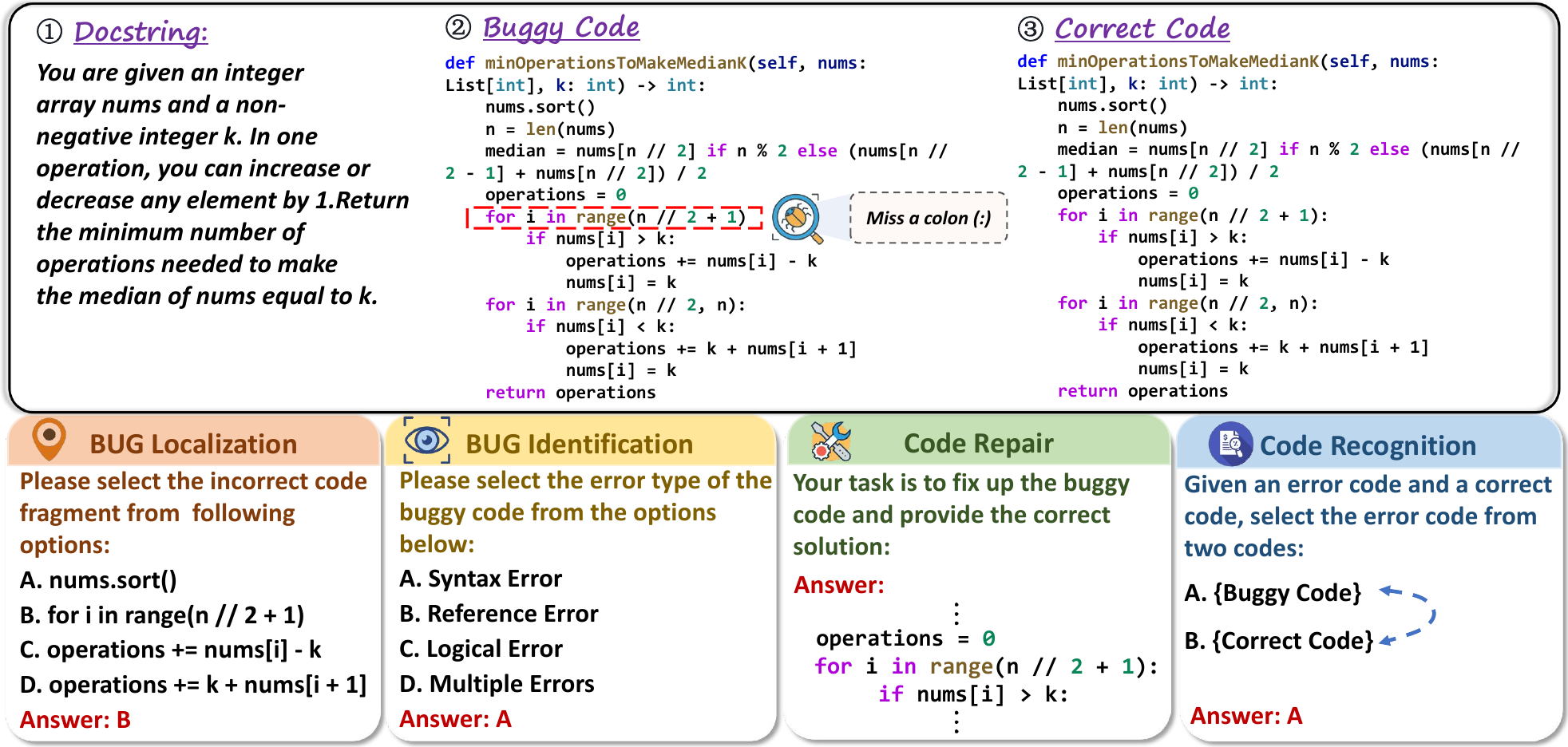}
    \caption{Illustration of \benchmark Benchmark. The \benchmark includes four key tasks: BUG Localization, BUG Identification, Code Repair, and Code Recognition.} \label{fig:benchmark}
\end{figure*}
Thrived on the emergent ability of Large Language Models (LLMs)~\cite{wei2022emergent}, recent research has increasingly focused on leveraging LLMs for automated debugging. Self-Debug~\cite{chen2023teaching} guides LLMs to generate code reviews, enabling them to refine their own buggy outputs. Self-Repair~\cite{olausson2023self} incorporates human feedback to address errors in generated code. Moreover, Self-Edit~\cite{zhang2023self} introduces a fault-aware editor that utilizes both error messages from test case evaluations and corresponding code snippets to repair bugs. \citet{wang2024intervenor} further explore the interactive Chain-of-Repair (CoR), which asks LLMs to iteratively refine code based on compiler error messages and generated repair guidelines. Despite these advances, the success of these methods heavily depends on the inherent debugging proficiency of LLMs.

To enhance the inherent debugging capabilities of LLMs, recent studies have emphasized generating data for supervised fine-tuning. InstructCoder~\cite{li2024instructcoderinstructiontuninglarge} adopts the Self-Instruct methodology~\cite{wang2023selfinstructaligninglanguagemodels} to construct an instruction-tuning dataset, which is used to improve LLMs' performance in debugging tasks. Similarly, \citet{li2024comprehensiveevaluationparameterefficientfinetuning} build the APR-INSTRUCTION dataset and finetune LLMs using four Parameter-Efficient Fine-Tuning (PEFT) techniques, including LoRA~\cite{J._Shen_Wallis_Allen-Zhu_Li_Wang_Chen_2021}, p-tuning~\cite{Li_Liang_2021}, prefix-tuning~\cite{Liu_Ji_Fu_Tam_Du_Yang_Tang_2021}, and $(\text{IA})^3$~\cite{Liu_Tam_Muqeeth_Mohta_Huang_Bansal_Raffel}. 

To evaluate the debugging capabilities of LLMs, several studies have developed comprehensive benchmarks. For example, \citet{wang2024intervenor} curate buggy code submissions from the Atcoder platform to create CodeError, a benchmark for evaluating the ability of LLMs to fix Python code. Similarly, \citet{tian2024debugbench} introduce DebugBench, which evaluates the debugging performance of LLMs in Python, C++, and Java by synthesizing buggy code using GPT-4~\cite{achiam2023gpt}. Lastly, \citet{guo2024codeeditorbench} also propose CodeEditorBench, a benchmark designed to evaluate the debugging capabilities of buggy code of varying difficulty levels across Python, Java, and C++. However, these benchmarks focus primarily on the Code Repair task and fail to provide a more holistic evaluation of LLMs' debugging abilities.

\section{\benchmark: Benchmarking the Debugging Capabilities of LLMs}

In this section, we first describe task definitions of designed tasks in \benchmark (Sec.~\ref{benchmark:def}). Then we detail the process of constructing the \benchmark benchmark (Sec.~\ref{benchmark:cons}). Finally, we show the characteristics of \benchmark by comparing with other different debugging benchmarks (Sec.~\ref{benchmark:compare}).

\begin{table*}[t]
\small
\centering
\caption{A Comparison between \benchmark and Other Code Debugging Benchmarks.}
\label{tab:comparision}
\resizebox{\linewidth}{!}{
\begin{tabular}{l|cccccc}
\hline
\textbf{Benchmark}           & \textbf{Number of Languages} & \textbf{Task}      & \textbf{Testing Set Scale} & \textbf{Error Types} & \textbf{Source of Bugs}             \\ \hline
DeepFix~\citeyearpar{yasunaga2021break}                    & 1                           & Code Repair        & 6,971                    & 4                    & User                                         \\
Review4Repair~\citeyearpar{huq2022review4repair}              & 1                           & Code Repair        & 2,961                    & -                    & User                                                \\
Bug2Fix~\citeyearpar{lu2021codexglue}                    & 1                           & Code Repair        & 5,835                    & -                    & User                                                \\
Github-Python~\citeyearpar{yasunaga2021break}              & 1                           & Code Repair        & 15,000                     & 14                   & User                                              \\
FixEval~\citeyearpar{haque2023fixeval}                    & 2                           & Code Repair        & 286,000                    & -                    & User                                                \\
CodeError~\citeyearpar{wang2024intervenor}                  & 1                           & Code Repair        & 4,463                    & 6                    & User                                              \\
xCodeEval~\citeyearpar{khan2023xcodeeval}                  & 11                          & Code Repair        & 17,699                   & 6                    & User                                               \\
DebugBench~\citeyearpar{tian2024debugbench}                 & 3                           & Code Repair        & 4,253                    & 18                   & GPT4                                             \\
CodeEditorBench~\citeyearpar{guo2024codeeditorbench}            & 3                           & Code Repair        & 1,907                    & 14                   & GPT4                                             \\ \hline
\multirow{4}{*}{\benchmark} & \multirow{4}{*}{3}          & BUG Localization   & \multirow{4}{*}{5,712}   & \multirow{4}{*}{18}  & \multirow{4}{*}{User\&GPT4}          \\
                           &                             & BUG Identification &                         &                      &                             &                            \\
                           &                             & Code Repair        &                         &                      &                             &                            \\
                           &                             & Code Recognition        &                         &                      &                             &                          \\ \hline
\end{tabular}
}
\end{table*}
\subsection{Task Definition}\label{benchmark:def}

As shown in Figure~\ref{fig:benchmark}, \benchmark introduces four distinct tasks to evaluate the debugging capabilities of LLMs. The tasks--BUG Localization, BUG Identification, and Code Recognition--are all single-choice problems, while the Code Repair task focuses on correcting buggy code.

\textbf{BUG Localization.} The BUG Localization task aims to identify the specific line of code that contains the bug. In this task, we prompt LLMs to select the line containing the bug from four given choices. Instead of asking LLMs to directly generate the buggy code line, we frame the Bug Localization task as a single-choice problem to ensure a more accurate evaluation.

\textbf{BUG Identification.} After pinpointing the code line that contains the bug, the software developers typically analyze the type of bug to facilitate more effective correction. For the BUG Identification task, LLMs are responsible for classifying the error type based on the provided buggy code. The possible error types include Syntax Error, Reference Error, Logical Error, and Multiple Errors.

\textbf{Code Repair.} The Code Repair~\cite{wang2024intervenor,tian2024debugbench} task requires LLMs to generate a corrected version of the given buggy code. The performance of LLMs in code repair is evaluated according to the correctness of the repaired code, and the correctness of the repaired code is evaluated by running a set of predefined test cases.

\textbf{Code Recognition.} The Code Recognition task provides two code segments and asks LLMs to identify which one contains the bug. Both correct and buggy codes differ in only a few lines.

\subsection{Details of Data Construction}\label{benchmark:cons}

To ensure the quality of \benchmark, we collect data from existing benchmarks (\texttt{DebugBench}~\cite{tian2024debugbench} and \texttt{LiveCodeBench}~\cite{jain2024livecodebench}) and human trials (\texttt{AtCoder} website\footnote{\url{https://atcoder.jp}}).

\textbf{Bug Localization.} In the Bug Localization task, we sample test instances from \texttt{DebugBench}. Then we construct the dataset by selecting no more than 20 instances for each of the 15 distinct single code error types across various programming languages. For each instance, we compare the buggy code with the corresponding correct code, identifying the line containing the bug as the golden reference. Additionally, we randomly select three other lines from the buggy code as distractors. Instances where fixing the bug requires inserting or deleting lines are excluded to streamline the construction of evaluation options for the LLMs. Each data entry is manually reviewed to eliminate weak or duplicate candidates, ensuring high quality and relevance.

\textbf{Bug Identification.} For the Bug Identification task, we sample test instances from \texttt{DebugBench}. The choices of this task consist of four error types: Syntax Error, Reference Error, Logical Error, and Multiple Errors. We compare the sizes of the test sample sets for each error type and select the smallest set size as the sampling number. The test instances are then sampled from the various code error sets based on this number.

\textbf{Code Repair.} For the Code Repair task, we first collect 138 programming contest problems newly published on \texttt{AtCoder} website between September 1, 2023, and April 1, 2024. We then gather buggy code submissions from users, specifically selecting examples in Python, C++, and Java. Additionally, we collect test cases for each problem to validate the correctness of the LLM-modified code.

\textbf{Code Recognition.} We build a mixed dataset by collecting test instances from both \texttt{DebugBench} and \texttt{LiveCodeBench}. For each programming language, we randomly select 800 test instances from this dataset. Each test instance consists of both a buggy code snippet and a correct version.

\subsection{Comparison of Different Debugging Benchmarks}\label{benchmark:compare}
Finally, we present a comprehensive comparison between \benchmark and other debugging benchmarks in Table~\ref{tab:comparision}, highlighting the characteristics and strengths of \benchmark.

Unlike existing debugging benchmarks, \benchmark offers a more comprehensive evaluation of debugging capabilities of LLMs by designing testing scenarios related to human-driven code debugging. To better simulate real-world software development, \benchmark includes buggy code samples generated both by humans and GPT-4. Furthermore, it spans a wide range of error types and different programming languages, more in line with the complexity of the real-world code debugging task. All these aspects ensure that \benchmark can provide more reliable evaluation results.

\section{COAST: Communicative Agent Based Data Synthesis Framework}
\begin{figure*}[t] \centering
    \includegraphics[width=1.0\textwidth]{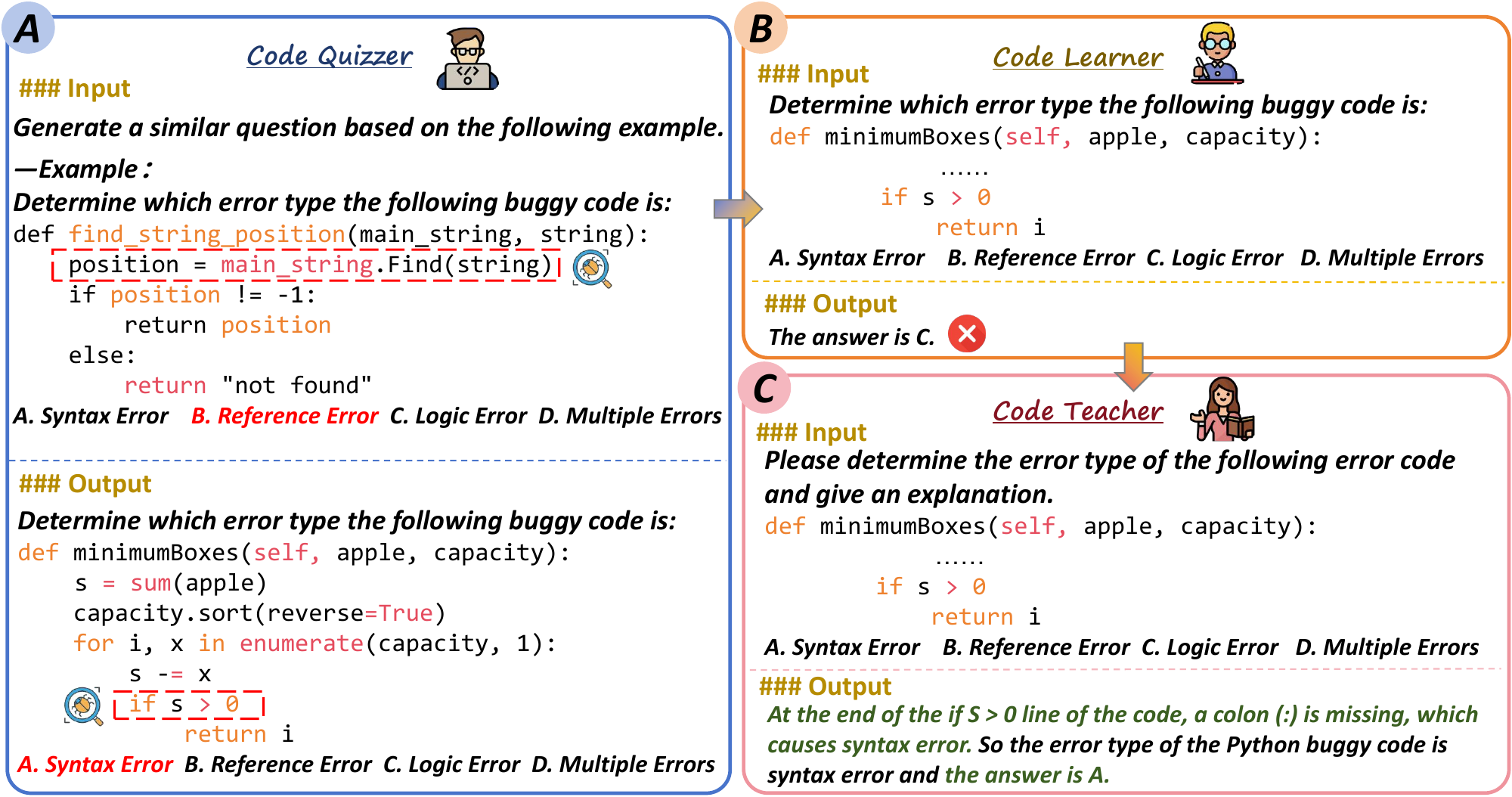}
    \caption{Illustration of \textbf{CO}mmunicative \textbf{A}gent Based Data \textbf{S}yn\textbf{T}hesis (COAST) Framework.} \label{fig:MASTER}
\end{figure*}

This section introduces the \textbf{CO}mmunicative \textbf{A}gent based data \textbf{S}yn\textbf{T}hesis (COAST) framework. As shown in Figure~\ref{fig:MASTER}, COAST automatically synthesizes high-quality debugging data through multi-agent interactions. We first configure the agents to play different roles in the COAST framework (Sec.~\ref{sec:model:agent}). Then COAST employs different agents to collaboratively synthesize high-quality training data for Supervised Fine-Tuning. (Sec.~\ref{sec:model:workflow}).


\subsection{Agent Building}
\label{sec:model:agent}
COAST employs three agents that collaboratively generate, resolve, and explain code debugging problems to synthesize high-quality SFT data. More details on the design of the prompts for different agents are provided in the Appendix~\ref{sec:appendix:prompt}. 

\textbf{Code Quizzer.} The \texttt{Code Quizzer} is designed to generate a set of code debugging problems. A stronger LLM is utilized as the \texttt{Code Quizzer}, which is configured using the instruction: ``You are a code debugging expert, skilled in generating code debugging problems to challenge programmers''. The \texttt{Code Quizzer} leverages examples from the \benchmark benchmark that span diverse debugging tasks and programming languages. These examples are then used to guide the \texttt{Code Quizzer} in generating corresponding task problems in various programming languages, which are subsequently solved by the \texttt{Code Learner}.

\textbf{Code Learner.} COAST framework sets \texttt{Code Learner}~\cite{lee2024llm2llm} to improve the quality of training data. The \texttt{Code Learner} uses the same backbone model as the SFT model and serves as a critic to evaluate the educational value of the problems generated by \texttt{Code Quizzer}. The \texttt{Code Learner} is instructed using the prompt: ``You are a code debugger''. Thus, \texttt{Code Learner} attempts to solve problems based on memorized knowledge and judge the educational value of each problem according to whether \texttt{Code Learner} successfully solves this problem. These educational instances can help improve the performance of \texttt{Code Learner} during Supervised Fine-Tuning.

\textbf{Code Teacher.} Inspired by~\citet{wang2024intervenor}, we also develop a \texttt{Code Teacher} by prompting the same LLM used for the \texttt{Code Quizzer} with the instruction: ``You are an experienced and insightful code debugger''. This prompt directs the LLM to act as a proficient code debugger, generating detailed explanations in the form of Chain-of-Thought (CoT)~\cite{wei2022chain}. These explanations provide valuable guidance to the \texttt{Code Learner} during the SFT process. 

\subsection{Synthesizing SFT Data through Multi-Agent Interactions}
\label{sec:model:workflow}
COAST automatically synthesizes high-quality data through the interaction among \texttt{Code Quizzer}, \texttt{Code Learner}, and \texttt{Coder Teacher}. The data synthesis process is detailed as follows.

In \textbf{Step A}, to ensure the diversity of tasks and error types during data synthesis, we instruct the \texttt{Code Quizzer} to generate debugging problems using demonstrations of different tasks and error types. Then, in \textbf{Step B}, the \texttt{Code Learner} acts as a critic, assessing the educational value of each synthesized problem. If the \texttt{Code Learner} solves the problem correctly, it indicates that the \texttt{Code Learner} already possesses the necessary knowledge to solve the problem and thus the problem can be discarded. On the other hand, if the \texttt{Code Learner} produces an incorrect solution, the problem is reserved as SFT data, due to its educational value for guiding the \texttt{Code Learner}. Finally, in \textbf{Step C}, the \texttt{Code Teacher} reviews the reserved problems and generates a detailed explanation for each problem. These explanations, in the form of Chain-of-Thought (CoT), include the analysis of errors and correct answers, which are essential for the \texttt{Code Learner} to understand the problems and refine their solutions. The responses generated by the \texttt{Code Teacher} are treated as the final outputs of the synthesized problems, forming the SFT data used to finetune \texttt{Code Learner}.

\begin{table}
\small
\centering
\caption{Statistics of Data Used in Different Supervised Fine-Tuning Strategies.}
\label{tab:training data quantity}
\begin{tabular}{l|l|r}
\hline
\textbf{SFT Data} & \textbf{Data Source} & \textbf{\#Instance} \\ \hline
\multirow{3}{*}{Human/GPT-4} & UltraInteract~\citeyearpar{yuan2024advancing} & 154,347 \\
 & InstructCoder~\citeyearpar{hu2023instructcoder} & 6,913 \\
 & RepairLlama~\citeyearpar{repairllama2023} & 64,643 \\ \hline
\multirow{4}{*}{COAST} & BUG Localization data & 4,681 \\
 & BUG Identification data & 4,474 \\
 & Code Repair data & 11,317 \\ 
 & Code Recognition data & 4,420 \\ \hline
 
\end{tabular}
\end{table}
\section{Experimental Methodology}
In this section, we describe the SFT dataset, evaluation metrics, evaluation models, and implementation details of our experiments.

\textbf{SFT Dataset.}
For Vanilla SFT, we collect training data from UltraInteract~\cite{yuan2024advancing}, InstructCoder~\cite{hu2023instructcoder}, and RepairLlama~\cite{repairllama2023} to finetune LLMs. These datasets are generated by GPT-4 or manually annotated by humans. For COAST, we construct training data for different tasks defined by \benchmark and the training data are mixed for training models. The data statistics are presented in Table~\ref{tab:training data quantity}. 

\textbf{Evaluation Metrics.} For the BUG Localization, BUG Identification, and Code Recognition tasks, LLMs are required to select an answer from multiple choices. Following prior work~\cite{suzgun2022challengingbigbenchtaskschainofthought}, we use Accuracy as the evaluation metric for these tasks. In particular, for the Code Recognition task, we swap the order of the two options, considering the problem is correctly solved only if LLMs answer correctly in both orders. For the Code Repair task, we adopt Pass@1~\cite{chen2021evaluating} to assess the effectiveness of various LLMs.

\textbf{Evaluation Models.}
We evaluate 13 LLMs on \benchmark, including both closed-source and open-source LLMs. Closed-source LLMs comprise the GPT series (GPT-4o-mini and GPT-3.5-Turbo). Open-source LLMs include the DeepSeek series (DeepSeek-Coder-V2, DeepSeek-V2, Deepseek-Coder-33B-Ins, DeepSeek-Coder-6.7B-Ins, and DeepSeek-LLM-7B-Ins), the Llama series (Llama3-70B-Ins, Llama3-8B-Ins, Llama2-7B-Ins, and CodeLlama-7B-Ins), and the Qwen series (Qwen2-72B-Ins and CodeQwen1.5-7B-Ins). More detailed descriptions of the evaluation models are shown in Appendix~\ref{sec:appendix:models}.

\textbf{Implementation Details.}
For closed-source LLMs, we utilize the APIs provided by their respective vendors. For open-source LLMs, we employ the vLLM framework~\cite{kwon2023efficient} for inference. During inference, we set the temperature to 0.2 and limit the maximum generation length to 1024 tokens. Our \texttt{Code Quizzer} and \texttt{Code Teacher} models are based on DeepSeek-Coder-V2, while DeepSeek-Coder-6.7B-Ins and Llama3-8B-Ins serve as our \texttt{Code Learner} models. All LLMs are trained using the Llama-Factory framework~\cite{llama-factory} using LoRA~\cite{hu2022lora} for efficient fine-tuning. For SFT, we configure the learning rate to 2e-5, set the number of training epochs to 1, use a batch size of 8, and employ 4 gradient accumulation steps.

\begin{table*}[t]
\small
\centering
\caption{Evaluation Results for Different LLMs on \benchmark, with DS representing the DeepSeek Models.}
\label{tab:Four Tasks}
\resizebox{\linewidth}{!}{
\begin{tabular}{l|cccc|cccc|cccc|cccc|c}
\hline
\multicolumn{1}{l|}{\multirow{2}{*}{\textbf{Model}}} & \multicolumn{4}{c|}{\textbf{BUG Localization}} & \multicolumn{4}{c|}{\textbf{BUG Identification}} & \multicolumn{4}{c|}{\textbf{Code Repair}} & \multicolumn{4}{c|}{\textbf{Code Recognition}} & \multicolumn{1}{c}{\multirow{2}{*}{\textbf{Avg.}}} \\ \cline{2-17}
\multicolumn{1}{c|}{} & \multicolumn{1}{l}{\textbf{PY}} & \multicolumn{1}{l}{\textbf{C++}} & \multicolumn{1}{l}{\textbf{JAVA}} & \multicolumn{1}{l|}{\textbf{Avg.}} & \multicolumn{1}{l}{\textbf{PY}} & \multicolumn{1}{l}{\textbf{C++}} & \multicolumn{1}{l}{\textbf{JAVA}} & \multicolumn{1}{l|}{\textbf{Avg.}} & \textbf{PY} & \textbf{C++} & \textbf{JAVA} & \textbf{Avg.} & \multicolumn{1}{l}{\textbf{PY}} & \multicolumn{1}{l}{\textbf{C++}} & \multicolumn{1}{l}{\textbf{JAVA}} & \multicolumn{1}{l|}{\textbf{Avg.}} & \multicolumn{1}{c}{} \\ \hline
GPT-4o-mini~\citeyearpar{gpt-4o-mini} & 84.8 & 81.0 & 81.5 & 82.4 & 53.3 & 48.5 & 48.9 & 50.2 & 65.2 & 67.2 & 67.4 & 66.6 & 85.4 & 90.9 & 91.0 & 89.1 & 72.1 \\
GPT-3.5-Turbo~\citeyearpar{chatgpt} & 40.4 & 47.2 & 52.2 & 46.9 & 35.5 & 33.3 & 34.1 & 34.3 & 57.2 & 52.9 & 61.6 & 57.2 & 79.4 & 82.4 & 84.0 & 81.9 & 55.1 \\
DeepSeek-V2~\citeyearpar{deepseekv2} & 82.0 & 81.0 & 85.9 & 83.0 & 62.0 & 61.0 & 61.3 & 61.4 & 65.2 & 63.0 & 63.5 & 63.9  & 77.4 & 83.9 & 80.5 & 80.6 & 72.2 \\
DeepSeek-Coder-V2~\citeyearpar{zhu2024deepseek} & 88.8 & 83.1 & 89.8 & 87.2 & 58.7 & 58.9 & 60.8 & 59.4 & 66.7 & 63.1 & 62.3 & 64.0 & 87.9 & 94.9 & 93.3 & 92.0 & 75.7 \\
Llama3-70B-Ins~\citeyearpar{llama3modelcard} & 74.2 & 75.9 & 82.0 & 77.5 & 42.8 & 42.3 & 44.9 & 43.3 & 44.9 & 44.2 & 45.7 & 44.9 & 73.9 & 61.6 & 63.3 & 66.3 & 58.0 \\
Qwen2-72B-Ins~\citeyearpar{qwen2} & 79.8 & 69.2 & 74.6 & 74.4 & 45.8 & 45.0 & 41.3 & 44.1 & 43.5 & 42.0 & 42.8 & 42.8 & 61.5 & 75.8 & 70.4 & 69.2 & 57.6 \\
DSCoder-33B-Ins~\citeyearpar{guo2024deepseekcoderlargelanguagemodel} & 52.2 & 50.3 & 51.7 & 51.4 & 24.9 & 26.0 & 30.9 & 27.2 & 46.4 & 50.7 & 54.3 & 50.5 & 24.8 & 27.0 & 30.5 & 27.4 & 39.1 \\
Llama2-7B-Ins~\citeyearpar{touvron2023llama} & 18.0 & 20.0 & 22.4 & 20.2 & 24.9 & 27.0 & 25.8 & 25.9 & 4.3 & 11.7 & 19.6 & 11.9 & 2.3 & 0.6 & 2.0 & 1.6 & 14.9 \\
CodeLlama-7B-Ins~\citeyearpar{roziere2023code} & 27.0 & 20.0 & 23.9 & 23.5 & 26.1 & 23.0 & 23.8 & 24.3 & 18.8 & 23.2 & 23.2 & 21.7 & 48.1 & 60.5 & 65.6 & 58.1 & 31.9 \\
CodeQwen1.5-7B-Ins~\citeyearpar{qwen} & 29.2 & 30.8 & 38.0 & 32.9 & 27.6 & 25.9 & 28.8 & 27.4 & 39.1 & 49.3 & 52.9 & 47.1  & 26.9 & 34.4 & 37.1 & 32.8 & 35.1 \\
DeepSeek-LLM-7B-Ins~\citeyearpar{deepseek-llm} & 27.0 & 19.0 & 25.9 & 23.9 & 30.5 & 28.5 & 30.9 & 29.9 & 21.0 & 24.1 & 14.5 & 19.9 & 35.9 & 36.6 & 46.0 & 39.5 & 28.3 \\ \hline
DSCoder-6.7B-Ins~\citeyearpar{guo2024deepseekcoderlargelanguagemodel} & 22.5 & 25.6 & 33.7 & 27.5 & 26.6 & 26.0 & 25.9 & 26.2 & 31.9 & 43.5 & 46.4 & 40.6 & 15.5 & 17.8 & 27.8 & 20.3 & 28.7 \\
NeuDebugger-DS-6.7B & 62.4 & 55.4 & 59.0 & 58.8 & 42.6 & 46.9 & 47.8 & 45.8 & 43.5 & 48.6 & 56.5 & 49.5 & 71.0 & 71.4 & 71.4 & 71.3 & 56.4 \\
Llama3-8B-Ins~\citeyearpar{llama3modelcard} & 55.6 & 55.9 & 61.0 & 57.6 & 36.8 & 38.1 & 34.6 & 36.6 & 26.1 & 34.3 & 28.3 & 29.6 & 69.1 & 77.4 & 78.1 & 74.9 & 49.7 \\
NeuDebugger-Llama3-8B & 64.6 & 57.9 & 61.0 & 61.1 & 38.6 & 29.9 & 33.3 & 33.8 & 38.4 & 41.3 & 45.7 & 41.8 & 75.3 & 78.0 & 82.4 & 78.5 & 53.8 \\ \hline
\end{tabular}
}
\end{table*}

\section{Evaluation Results}
In this section, we first show the performance of various LLMs on \benchmark and also conduct ablation studies to investigate the effectiveness of NeuDebugger models. Then the effectiveness of NeuDebugger models is evaluated in addressing different types of code errors. Finally, some case studies are provided in the Appendix~\ref{sec:appendix:case_studies}.


\begin{table}
\small
\centering
\caption{Effectiveness of Different SFT Strategies.}
\label{tab:Analysis 1}
\resizebox{\linewidth}{!}{
\begin{tabular}{llllll}
\hline
\multicolumn{1}{l|}{{\textbf{Method}}} & \textbf{\begin{tabular}[c]{@{}c@{}}BUG \\ Loc.\end{tabular}} & \textbf{\begin{tabular}[c]{@{}c@{}}BUG \\ Iden.\end{tabular}} & \textbf{\begin{tabular}[c]{@{}c@{}}Code \\ Rep.\end{tabular}} & \textbf{\begin{tabular}[c]{@{}c@{}}Code \\ Rec.\end{tabular}} & \textbf{Avg.} \\ \hline
\multicolumn{6}{c}{\textit{DSCoder-6.7B-Ins}} \\ \hline
\multicolumn{1}{l|}{Zero-Shot} & 27.5 & 26.2 & 40.6 & 20.3 & 28.7 \\
\multicolumn{1}{l|}{w/ Vanilla SFT} & 21.8 & 23.1 & 40.1 & 9.4 & 23.6 \\
\multicolumn{1}{l|}{w/ COAST (Answer)} & 43.8 & 35.8 & 43.5 & 32.7 & 39.0 \\
\multicolumn{1}{l|}{w/ COAST (CoT)} & \textbf{60.7} & 45.0 & 38.7 & 34.7 & 44.8 \\
\multicolumn{1}{l|}{NeuDebugger} & 58.8 & \textbf{45.8} & \textbf{49.5} & \textbf{71.3} & \textbf{56.4} \\ \hline
\multicolumn{6}{c}{\textit{Llama3-8B-Ins}} \\ \hline
\multicolumn{1}{l|}{Zero-Shot} & 57.6 & \textbf{36.6} & 29.6 & 74.9  & 49.7 \\
\multicolumn{1}{l|}{w/ Vanilla SFT} & 53.6 & 34.0 & 28.7 & 26.0 & 35.6 \\
\multicolumn{1}{l|}{w/ COAST (Answer)} & 58.1 & 34.8 & \textbf{42.5} & 32.1 & 41.9 \\
\multicolumn{1}{l|}{w/ COAST (CoT)} & \textbf{64.4} & 34.6 & 32.6 & 78.1 & 52.4 \\
\multicolumn{1}{l|}{NeuDebugger} & 61.1 & 33.8 & 41.8 & \textbf{78.5} & \textbf{53.8} \\ \hline
\end{tabular}
}
\end{table}
\begin{table*}
\small
\centering
\caption{Performance of Vanilla LLMs and NeuDebugger on Different Bug Types. Ref. denotes Reference and Multi. denotes Multiple.}
\label{tab:diff_bug_types}
\resizebox{\linewidth}{!}{
\begin{tabular}{l|l|cccc|cccc|cccc}
\hline
\multirow{2}{*}{\textbf{Task}} & \multirow{2}{*}{\textbf{Model}} & \multicolumn{4}{c|}{\textbf{Python}}                          & \multicolumn{4}{c|}{\textbf{C++}}                              & \multicolumn{4}{c}{\textbf{Java}}                              \\ \cline{3-14} 
                               &                                 & \textbf{Syntax}        & \textbf{Ref.}     & \textbf{Logic}       & \textbf{Multi.}      & \textbf{Syntax}        & \textbf{Ref.}     & \textbf{Logic}        & \textbf{Multi.}      & \textbf{Syntax}        & \textbf{Ref.}     & \textbf{Logic}        & \textbf{Multi.}      \\ \hline
\multirow{4}{*}{Bug Iden.}     & Llama3-8B-Ins                   & 0.0           & 3.7           & 97.9 & 45.8          & 0.0           & 3.0  & 87.5  & 62.0          & 0.0           & 3.7           & 87.4  & 47.4          \\
                               & NeuDebugger-Llama3-8B           & 34.2 & 16.8 & 27.9          & 75.3 & 17.0 & 3.0  & 15.0           & 84.5 & 25.3 & 4.2  & 23.2           & 80.5 \\
                               & DSCoder-6.7B-Ins                & 3.2           & 3.2           & 99.5 & 0.5           & 2.0           & 0.0           & 100.0 & 2.0           & 1.6           & 1.6           & 100.0 & 0.5           \\
                               & NeuDebugger-DS-6.7B             & 54.2 & 81.1 & 32.6          & 2.6  & 45.0 & 45.0 & 77.5           & 20.0 & 50.0 & 55.8 & 72.6           & 12.6 \\ \hline
\multirow{4}{*}{Code Rep.}     & Llama3-8B-Ins                   & 40.0          & 20.0          & 35.6          & 7.5           & 60.9          & 45.8          & 29.6           & 16.7 & 40.0          & 52.4          & 28.3           & 7.7           \\
                               & NeuDebugger-Llama3-8B           & 90.0 & 46.7 & 39.7 & 20.0 & 65.2 & 66.7 & 40.7  & 10.8          & 80.0 & 76.2 & 39.6  & 15.4 \\
                               & DSCoder-6.7B-Ins                & 40.0          & 33.3          & 38.4          & 17.5          & 69.6          & 58.3          & 40.7  & 21.6          & 60.0          & 71.4          & 47.2           & 23.1          \\
                               & NeuDebugger-DS-6.7B             & 90.0 & 46.7 & 45.2 & 27.5 & 87.0 & 62.5 & 40.7  & 27.0 & 76.0 & 85.7 & 54.7  & 30.8 \\ \hline
\end{tabular}
}
\end{table*}

\subsection{Overall Performance}

As shown in Table~\ref{tab:Four Tasks}, we show the debugging performance of different LLMs on \benchmark. 

Among the four tasks defined in \benchmark, LLMs typically excel in Bug Localization and Code Recognition tasks compared to other tasks, which demonstrates that LLMs have strong capabilities in code differentiation and bug detection. In contrast, these LLMs demonstrate lower effectiveness in both BUG Identification and Code Repair tasks. The poor performance of LLMs in the Bug Identification task highlights that current LLMs are weak at analyzing errors and cannot understand them effectively. The suboptimal performance of these 70B-scale LLMs in the Code Repair task further highlights the challenges they face in self-debugging~\cite{chen2023teaching}. This phenomenon has also been observed by~\citet{wang2024intervenor}.

The experimental results demonstrate that larger-scale LLMs generally exhibit superior code debugging capabilities compared to their smaller-scale counterparts. This illustrates the critical role of model scale in preserving emergent ability~\cite{wei2022emergent} and highlights the effectiveness of training on code-related data for acquiring specialized knowledge~\cite{yuan2024advancing,gudibande2024false}. Furthermore, we evaluate the performance of LLMs across various programming languages to investigate their effectiveness and robustness in addressing debugging challenges. Notably, some LLMs, such as GPT-4o-mini and DeepSeek-Coder-V2, show consistent performance across all programming languages, whereas some 7B-scale LLMs show significant variability in their performance across different languages. These observations highlight the necessity of establishing a comprehensive benchmark encompassing multiple programming languages to rigorously assess the debugging capabilities of LLMs.

Finally, we present the performance of NeuDebugger, which are finetuned using the synthesized data of COAST. These NeuDebugger models demonstrate superior code debugging capabilities compared to other 7B-scale LLMs, achieving performance on par with GPT-3.5-Turbo. Notably, both NeuDebugger-DS-6.7B and NeuDebugger-Llama3-8B outperform their corresponding foundation models across the four tasks outlined in \benchmark, with improvements of 27.7\% and 4.1\%, respectively. The phenomenon highlights the effectiveness of COAST in generating high-quality data to improve the debugging performance of LLMs.

\subsection{Ablation Studies}
As shown in Table~\ref{tab:Analysis 1}, we conduct ablation studies to evaluate the effectiveness of various SFT strategies, including Vanilla SFT, COAST (Answer), COAST (CoT), and NeuDebugger. Additionally, the impact of data quantity on the performance of the NeuDebugger strategy across different LLMs is further discussed in the Appendix~\ref{app:data_quantity}.

The Vanilla SFT strategy collects SFT data from UltraInteract~\cite{yuan2024advancing}, InstructCoder~\cite{hu2023instructcoder}, and RepairLlama~\cite{repairllama2023} to finetune LLMs. The COAST framework generates SFT data through multi-agent interactions and we conduct three different SFT strategies to show the effectiveness of different agents, including COAST (Answer), COAST (CoT), and NeuDebugger. Specifically, the COAST (Answer) model removes \texttt{Code Teacher} from COAST and directly takes correct answers as outputs for the synthesized data. COAST (CoT) asks the \texttt{Code Teacher} to generate reasoning process (thoughts) for problem-solving as the output. The NeuDebugger strategy combines both COAST (Answer) and COAST (CoT) strategies by merging training data from both strategies to finetune the \texttt{Code Learner}. Specifically, the data for the BUG Localization, BUG Identification, and Code Recognition tasks is sourced from COAST (CoT) strategy, while the Code Repair data come from COAST (Answer) strategy.

The evaluation results reveal that the Vanilla SFT strategy performs significantly worse than the baseline models, illustrating the challenge of ensuring high-quality data for finetuning LLMs. In contrast, using synthesized data generated by COAST markedly enhances the code debugging capabilities of these models. This suggests that both DSCoder-6.7B-Ins and Llama3-8B-Ins struggle to effectively acquire debugging knowledge from human or GPT-4 annotated data~\cite{gudibande2024false}. Moreover, the COAST (CoT) strategy consistently outperforms COAST (Answer) in most tasks, except for the Code Repair task. This exception might be due to the Chain-of-Thought (CoT) explanations produced by \texttt{Code Teacher}, which offer clearer reasoning but potentially introduce noise specific to Code Repair scenarios. By integrating SFT data from both COAST (CoT) and COAST (Answer), the NeuDebugger strategy achieves the best overall performance among all SFT methods. Collectively, these experimental results highlight the effectiveness of the COAST framework, which leverages multiple agents to synthesize high-quality SFT data for fine-tuning LLMs.

\subsection{Effectiveness of NeuDebugger on Different Bug Types}
In Table~\ref{tab:diff_bug_types}, we evaluate the effectiveness of NeuDebugger models in different bug types.

\begin{figure}[t]
    \centering
    \subfigure[DSCoder-6.7B-Ins.] { \label{fig:DSCoder}
    \includegraphics[width=0.48\linewidth]{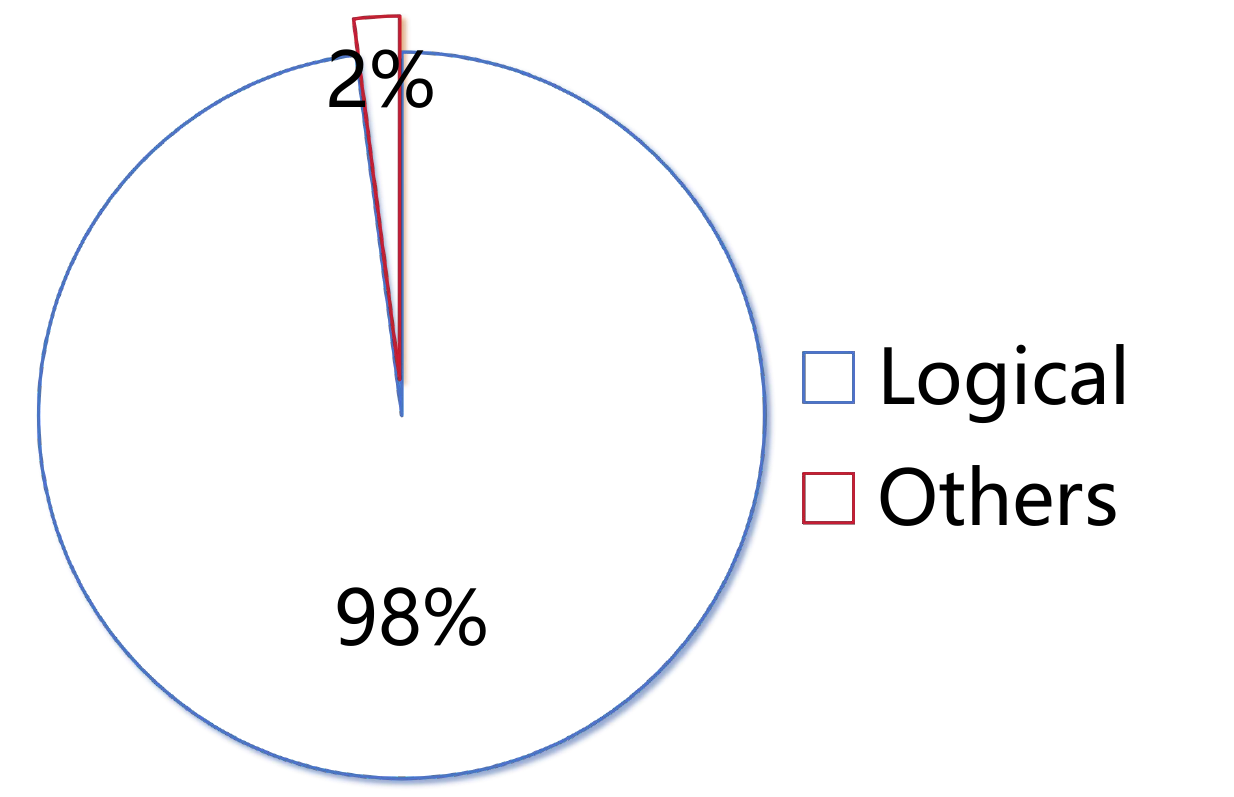}}
    \subfigure[NeuDebugger-DS-6.7B.] { \label{fig:NeuDS}
    \includegraphics[width=0.48\linewidth]{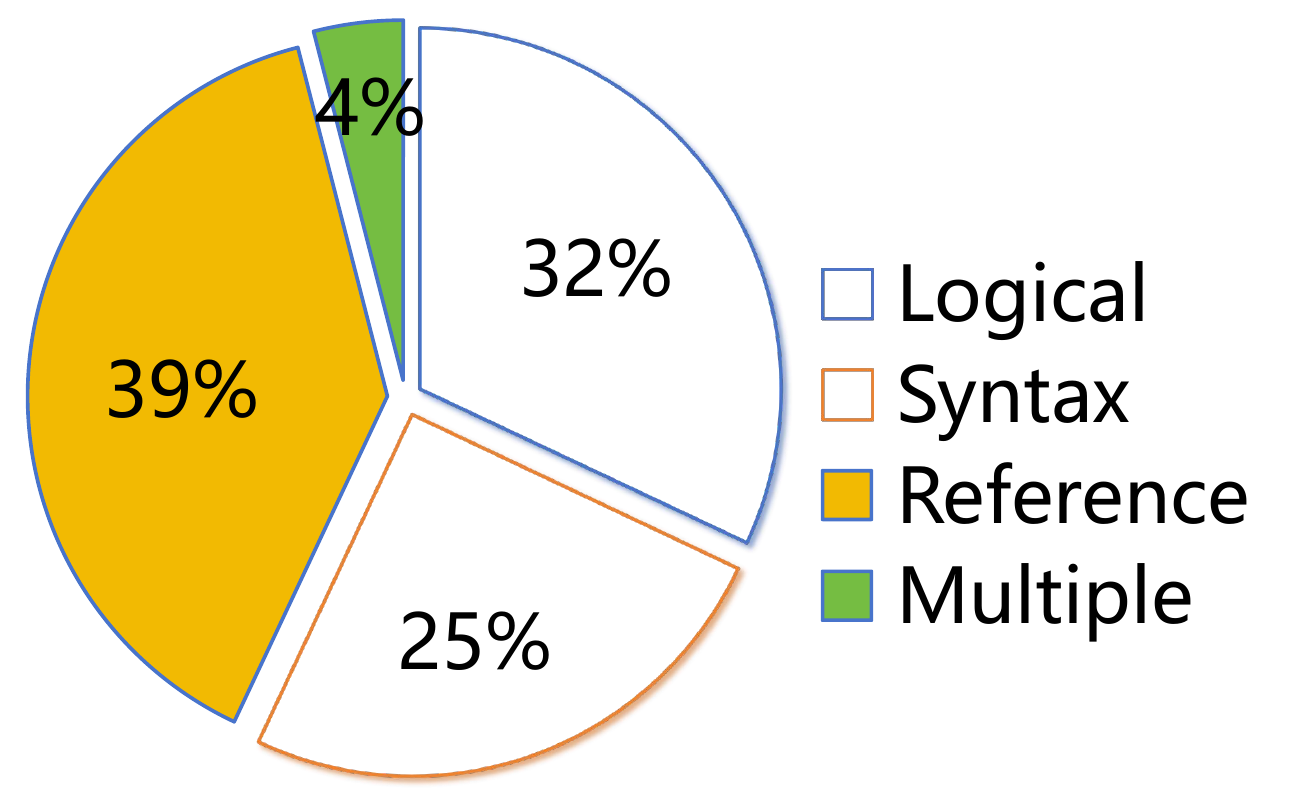}}
    \caption{Response Distributions of DSCoder-6.7B-Ins and NeuDebugger-DS-6.7B in BUG Identification Task.}
    \label{fig:Analysis 3}
\end{figure}

For the Bug Identification task, the evaluation results reveal that existing LLMs remain suboptimal in analyzing the root causes of bugs. Figure~\ref{fig:Analysis 3} further shows the response distributions of these models, which illustrates a notable limitation of DSCoder-6.7B-Ins. DSCoder-6.7B-Ins often defaults to selecting ``Logical Error'', leading to a relatively high accuracy for this category but a lack of nuanced identification among different bug types. In contrast, NeuDebugger-DS-6.7B showcases its effectiveness by achieving a more balanced distribution in option selection, resulting in significant improvements in overall Bug Identification performance. Additional results of other LLMs on this task are provided in the Appendix~\ref{sec:appendix:analysis}.

For the Code Repair task, NeuDebugger models achieve improvements on all bug types, particularly for Syntax Errors, showing the effectiveness of NeuDebugger models. This finding also suggests that Syntax Errors are relatively simpler and easier for LLMs to learn compared to the other bug types. However, NeuDebugger models still face challenges when addressing more complex issues such as Logic Errors and Multiple Errors.

\section{Conclusion}
This paper presents \benchmark, a comprehensive benchmark designed to evaluate the debugging capabilities of LLMs from multiple aspects. Our experiments reveal that 7B-scale LLMs consistently underperform in a wide range of debugging tasks. To address this limitation, we introduce COAST, a multi-agent based framework for synthesizing high-quality data to improve the debugging abilities of LLMs. 

\section*{Limitations}
While \benchmark includes four evaluation tasks, incorporating additional tasks could enhance its ability to better evaluate the debugging capabilities of LLMs. In our experiments, the COAST framework demonstrates strong potential in generating high-quality SFT data for LLMs. However, its effectiveness is constrained by the performance of the \texttt{Code Quizzer} and \texttt{Code Teacher} models. In particular, the quality of the generated data heavily relies on the capabilities and diversity of the foundation model of \texttt{Code Quizzer} and \texttt{Code Teacher}.

\section*{Acknowledgments}
This work is partly supported by the Natural Science Foundation of China under Grant (No. 62206042, No. 62137001, and No. 62272093), CCF-Zhipu Large Model Innovation Fund (No. 202403), the Joint Funds of Natural Science Foundation of Liaoning Province (No. 2023-MSBA-081), and the Fundamental Research Funds for the Central Universities under Grant (No. N2416012).



\bibliography{custom_1}
\clearpage
\newpage
\appendix

\section{Appendix}
\label{sec:appendix}

\subsection{License}
\label{license}
For all datasets in our experiments, DebugBench
uses the Apache License 2.0, LiveCodeBench uses the MIT License. All of these licenses allow their data to be used
for academic use.

\subsection{Prompt Templates of Different Agents Defined in COAST}
\label{sec:appendix:prompt}
\begin{figure}[t] \centering
    \includegraphics[width=1.0\linewidth]{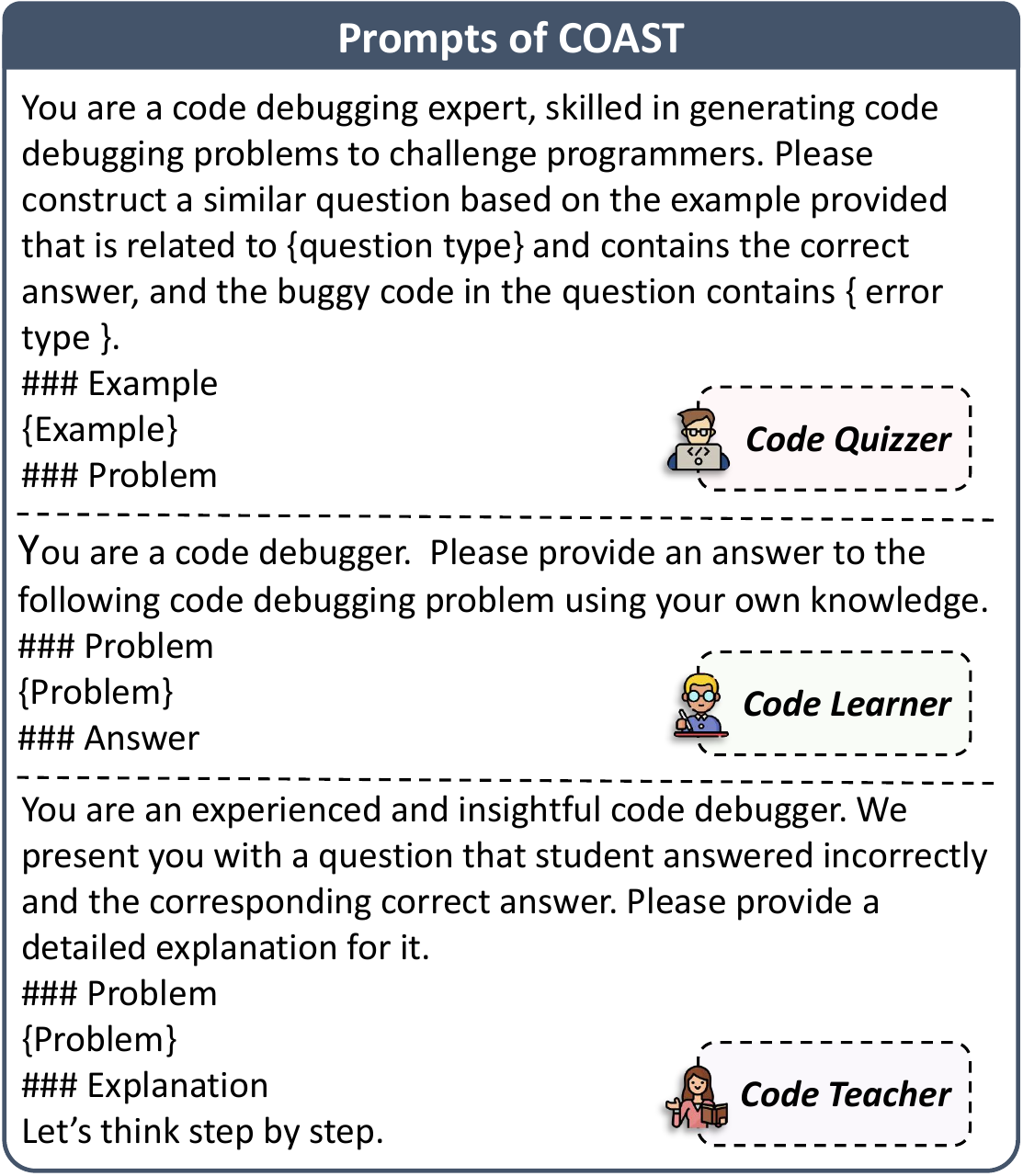}
    \caption{Illustrations of Prompts Used in COAST to Configure Different Agents. Within COAST, there are three LLM-based agents, including \texttt{Code Quizzer}, \texttt{Code Learner}, and \texttt{Code Teacher}. We utilize specific instructions to ensure they play the correct roles and carry out the intended tasks.} \label{fig:prompt}
\end{figure}

As illustrated in Figure~\ref{fig:prompt}, we utilize different prompts to guide LLMs to play the roles of \texttt{Code Quizzer}, \texttt{Code Learner}, and \texttt{Code Teacher}. 

For each task defined in \benchmark, we individually select an instance for each programming language and then use them to prompt \texttt{Code Quizzer} to generate code debugging problems of different tasks, bug types, and programming languages. These problems are intended to evaluate \texttt{Code Learner}'s code debugging abilities. For the questions that \texttt{Code Learner} answers incorrectly, we use the following instruction: ``You are an experienced and insightful code debugger'' to prompt \texttt{Code Teacher} to generate explanations and solutions of the questions. Finally, questions, solutions, and explanations are collected to construct the SFT data for fine-tuning \texttt{Code Learner}.

\subsection{Details of LLMs Used for Evaluation}
\label{sec:appendix:models}
In this subsection, we present a detailed description of the models evaluated on \benchmark.

\textit{OpenAI GPTs.} \texttt{GPT-4o-mini}~\cite{gpt-4o-mini} and \texttt{GPT-3.5-Turbo}~\cite{chatgpt} are two popular and powerful LLMs, which belong to different variants of the GPT family, developed by OpenAI. \texttt{GPT-4o-mini} is a lightweight version of the \texttt{GPT-4o} model, but still inherits the core advantages of the \texttt{GPT-4o}, including powerful text generation, logical reasoning, and code generation. Both models are black-box models, which supply commercial APIs for usage.

\textit{Meta Llama.} \texttt{Llama2-7B-Ins}~\cite{touvron2023llama} is an open-sourced LLM. It is trained with up to 1.4 trillion tokens, where 4.5\% of them are code tokens from Github. \texttt{CodeLlama-7B-Ins}~\cite{roziere2023code} conducts an additional instruction-tuning stage to adapt \texttt{Llama2}~\cite{touvron2023llama} to improve the effectiveness in code-related tasks. Recently, \texttt{Llama3}~\cite{llama3modelcard} models have been released, which is a major leap over \texttt{Llama2} models and establishes a new state-of-the-art.

\textit{Aliyun Qwen.} \texttt{Qwen2-72B-Ins} is a 72 billion parameter scale LLM. \texttt{Qwen2-72B} employs a variety of automated methods for obtaining high-quality instruction and preference data, making it perform well on code and maths tasks. \texttt{CodeQwen1.5-7B-Ins}~\cite{qwen} is the code-oriented version of \texttt{Qwen1.5-7B}~\cite{qwen1.5}. \texttt{CodeQwen1.5-7B-Ins} has been tuned with around 3 trillion tokens of code-related data. It supports 92 programming languages and supports long context understanding and generation with a context length of 64K tokens.

\textit{DeepSeek.} DeepSeek series models are released by High-Flyer. \texttt{DeepSeek-LLM-7B}~\cite{deepseek-llm} is trained from scratch with 2 trillion tokens in both English and Chinese. \texttt{DeepSeek-LLM-7B-Ins}~\cite{deepseek-llm} is initialized by \texttt{DeepSeek-LLM-7B} and tuned with an additional 1 million instruction data. 
\texttt{DSCoder-6.7B-Ins}~\cite{guo2024deepseekcoderlargelanguagemodel} and \texttt{DSCoder-33B-Ins}~\cite{guo2024deepseekcoderlargelanguagemodel} are trained from scratch on 2T tokens, which consist of 87\% code and 13\% natural language. \texttt{DeepSeek-V2}~\cite{deepseekv2} contains 236B parameters and employs the Mixture-of-Experts (MoE)~\cite{cai2024surveymixtureexperts} architecture to conduct efficient training and inference. It is trained on a high-quality corpus comprising 8.1 trillion tokens. \texttt{DeepSeek-Coder-V2}~\cite{zhu2024deepseek} is also an open-sourced MoE-based LLM, which achieves comparable performance with \texttt{GPT4-Turbo} in code-related tasks. \texttt{DeepSeek-Coder-V2} starts from an intermediate checkpoint of \texttt{DeepSeek-V2} and is tuned using 6 trillion tokens. 
\begin{figure}[t] \centering
    \includegraphics[width=0.5\textwidth]{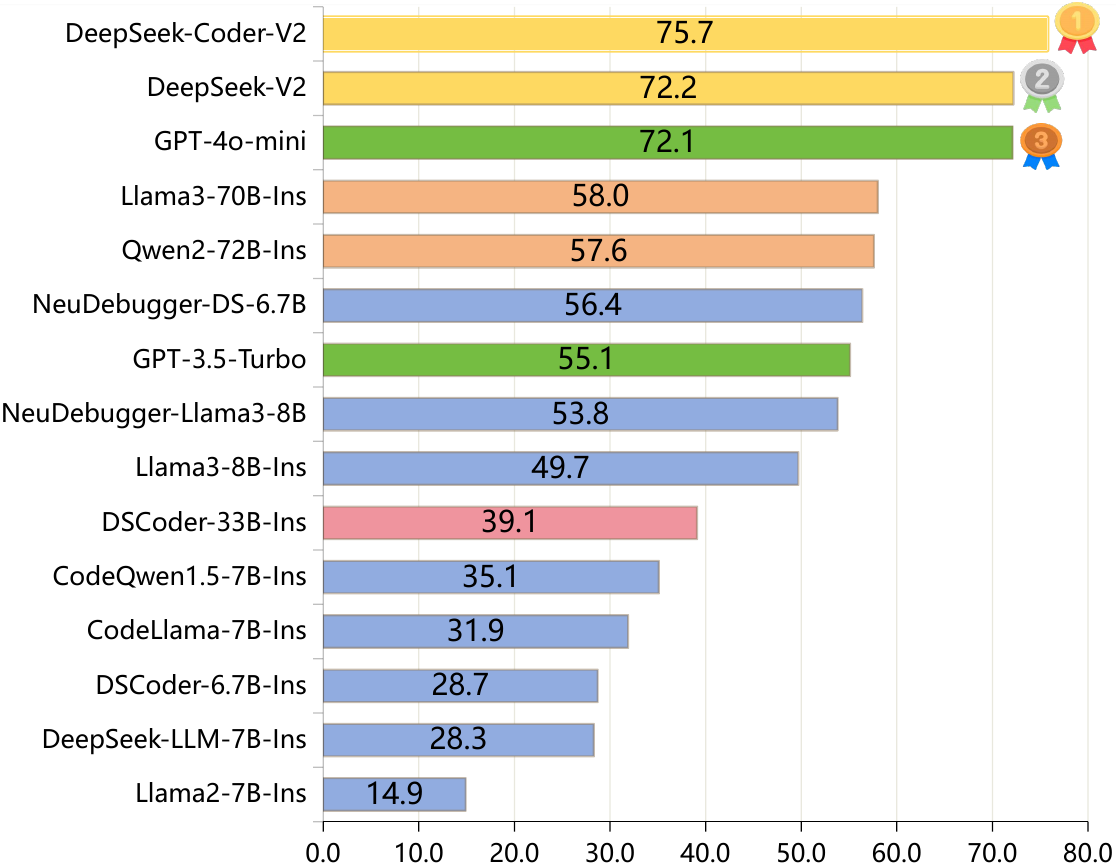}
    \caption{Debugging Performance of Different LLMs on \benchmark. Different LLMs are ranked according to their average performance. Different colors represent different LLM scales.\label{fig: ranking}}  
\end{figure}

\subsection{Performance Ranking on \benchmark}
As shown in Figure~\ref{fig: ranking}, we show the debugging performance of different LLMs on \benchmark.

The evaluation results show that larger-scale LLMs consistently outperform smaller-scale LLMs on \benchmark, highlighting the superior capacity of larger-scale LLMs to handle code debugging tasks. The DeepSeek series models exhibit particularly strong performance, with the open-source DeepSeek-Coder-V2 surpassing the closed-source GPT-4o-mini. NeuDebugger-DS-6.7B and NeuDebugger-Llama3-8B, based on DeepSeek-Coder-6.7B-Ins and Llama3-8B-Ins, respectively, demonstrate improvements of 27.7\% and 4.1\% when trained using the data synthesized by the COAST framework. It shows the effectiveness of COAST in generating high-quality training data to improve the debugging performance of LLMs.

\begin{figure}[t]
    \centering
    \subfigure[BUG Localization.] { \label{fig:bug loc}
    \includegraphics[width=0.48\linewidth]{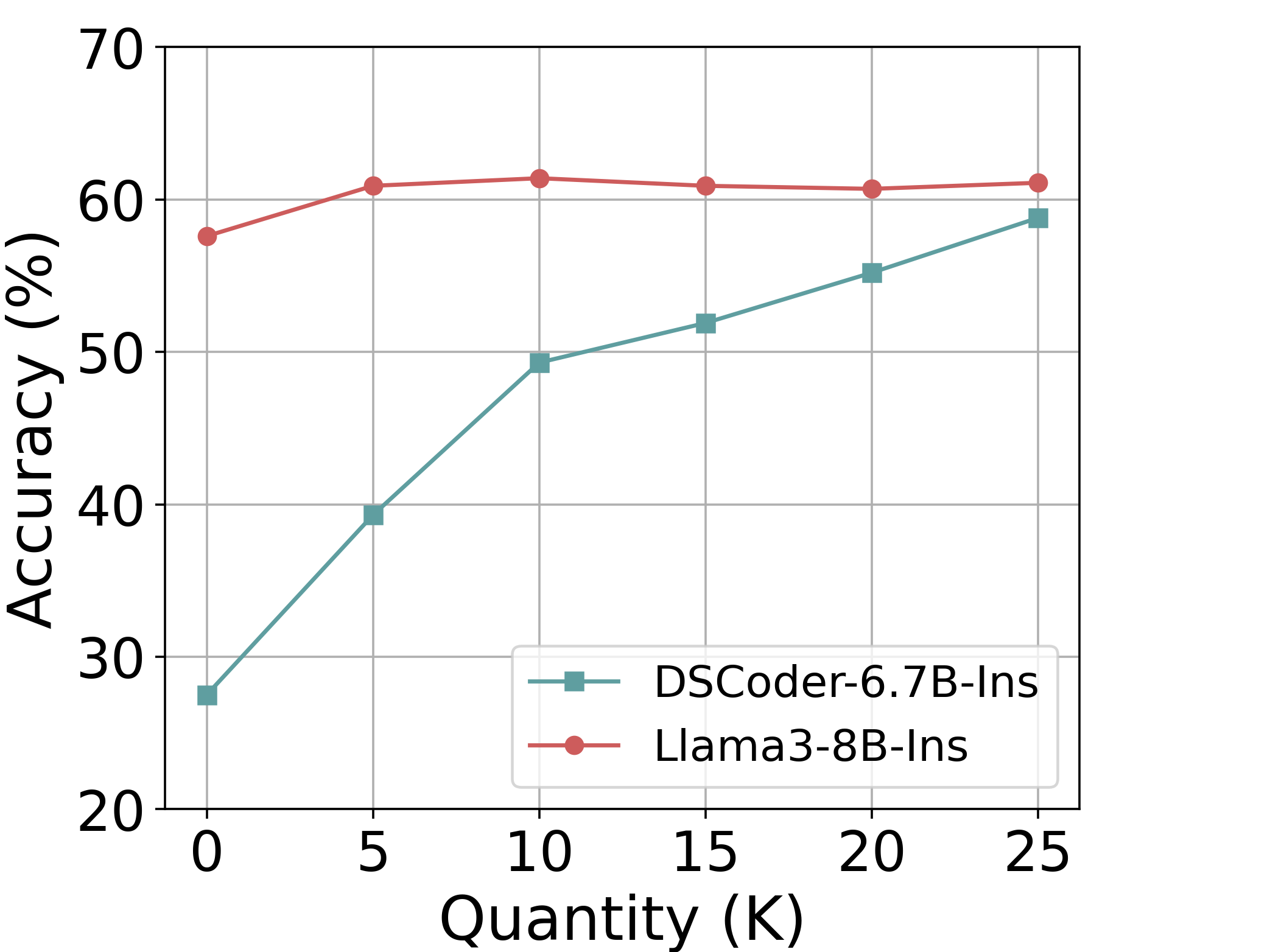}}
    \subfigure[BUG Identification.] { \label{fig:bug iden}
    \includegraphics[width=0.48\linewidth]{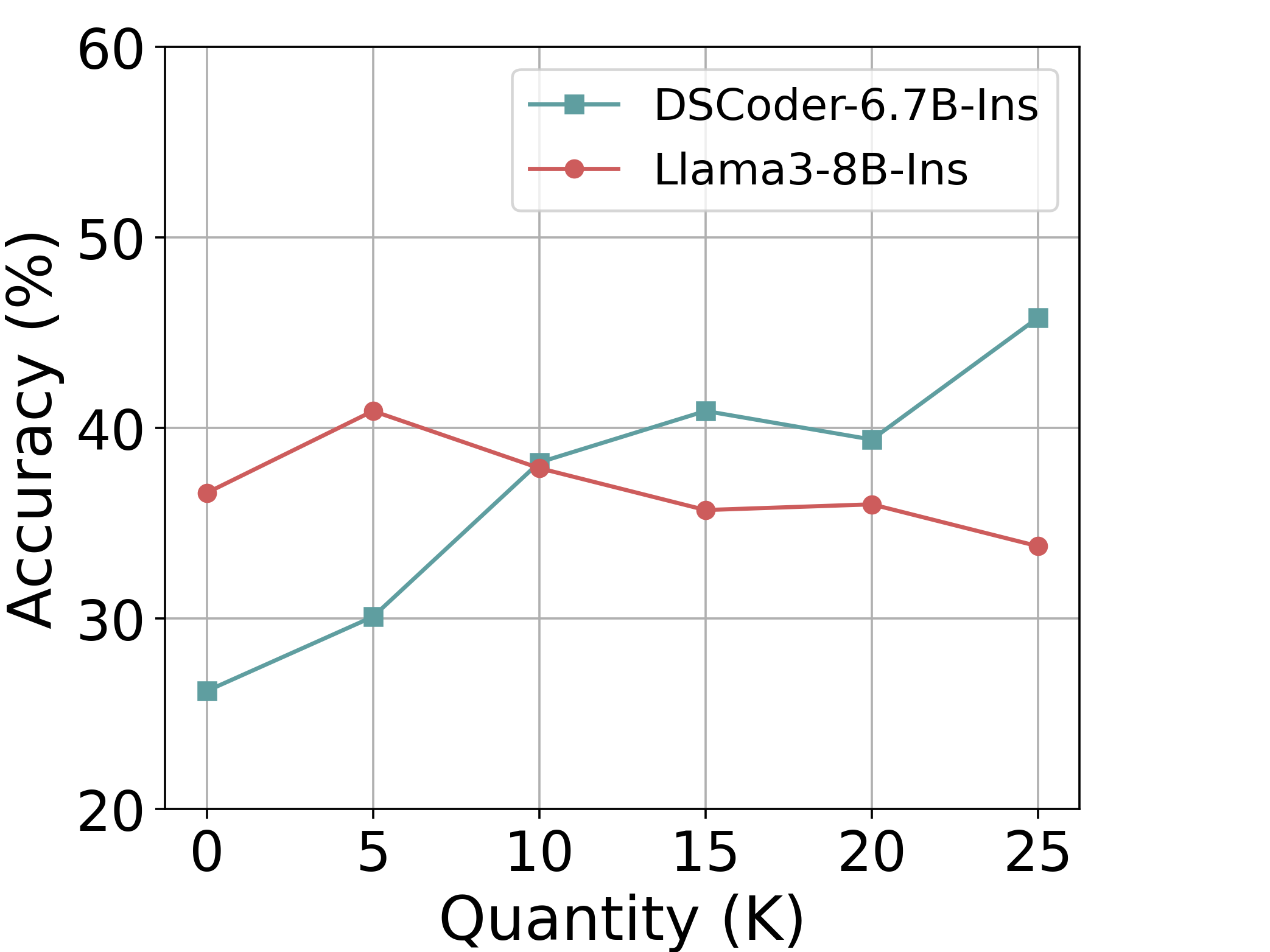}}
    \subfigure[Code Repair.] { \label{fig:code rep}
    \includegraphics[width=0.48\linewidth]{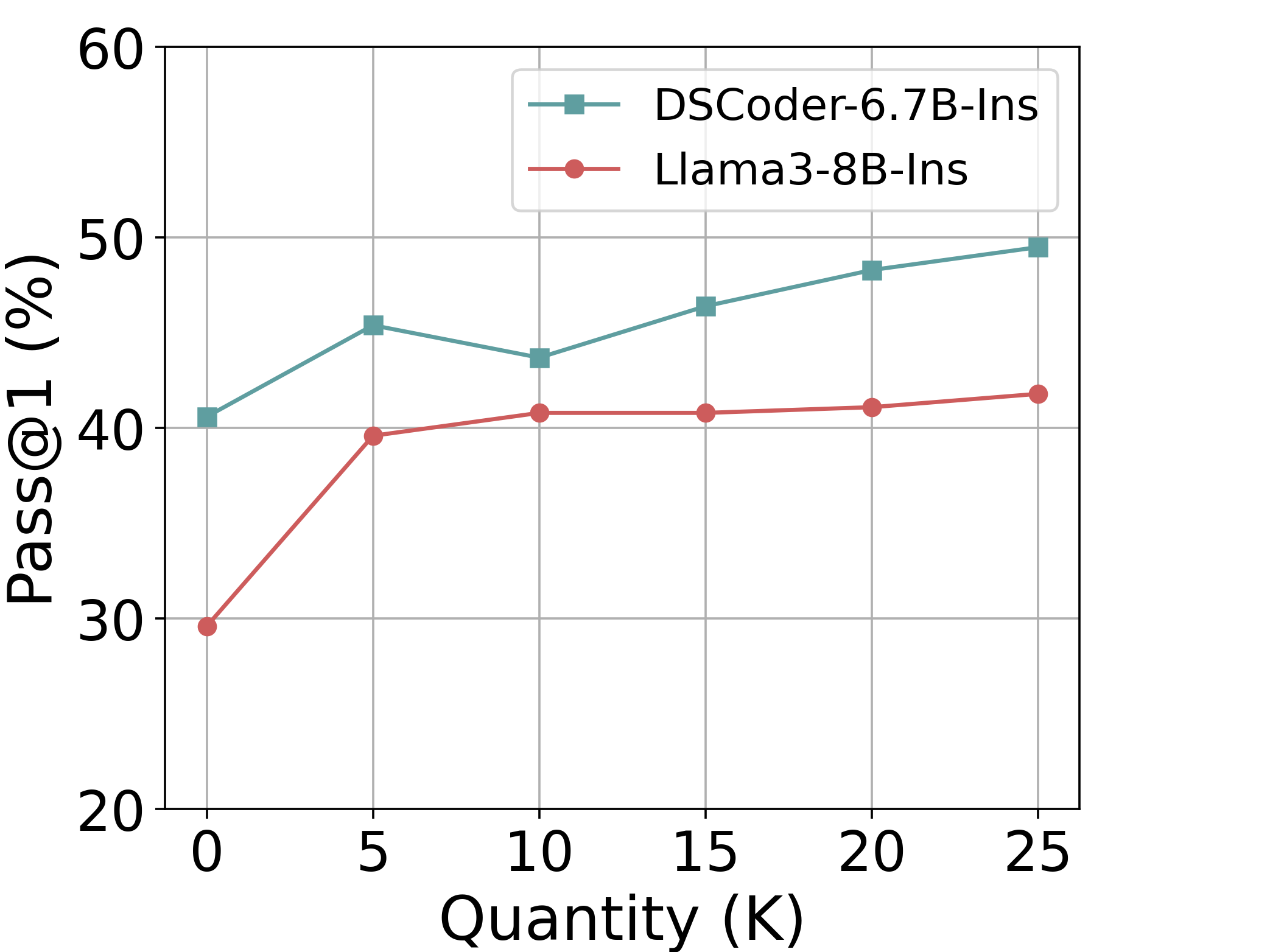}}
    \subfigure[Code Recognition.] { \label{fig:code rev} 
    \includegraphics[width=0.48\linewidth]{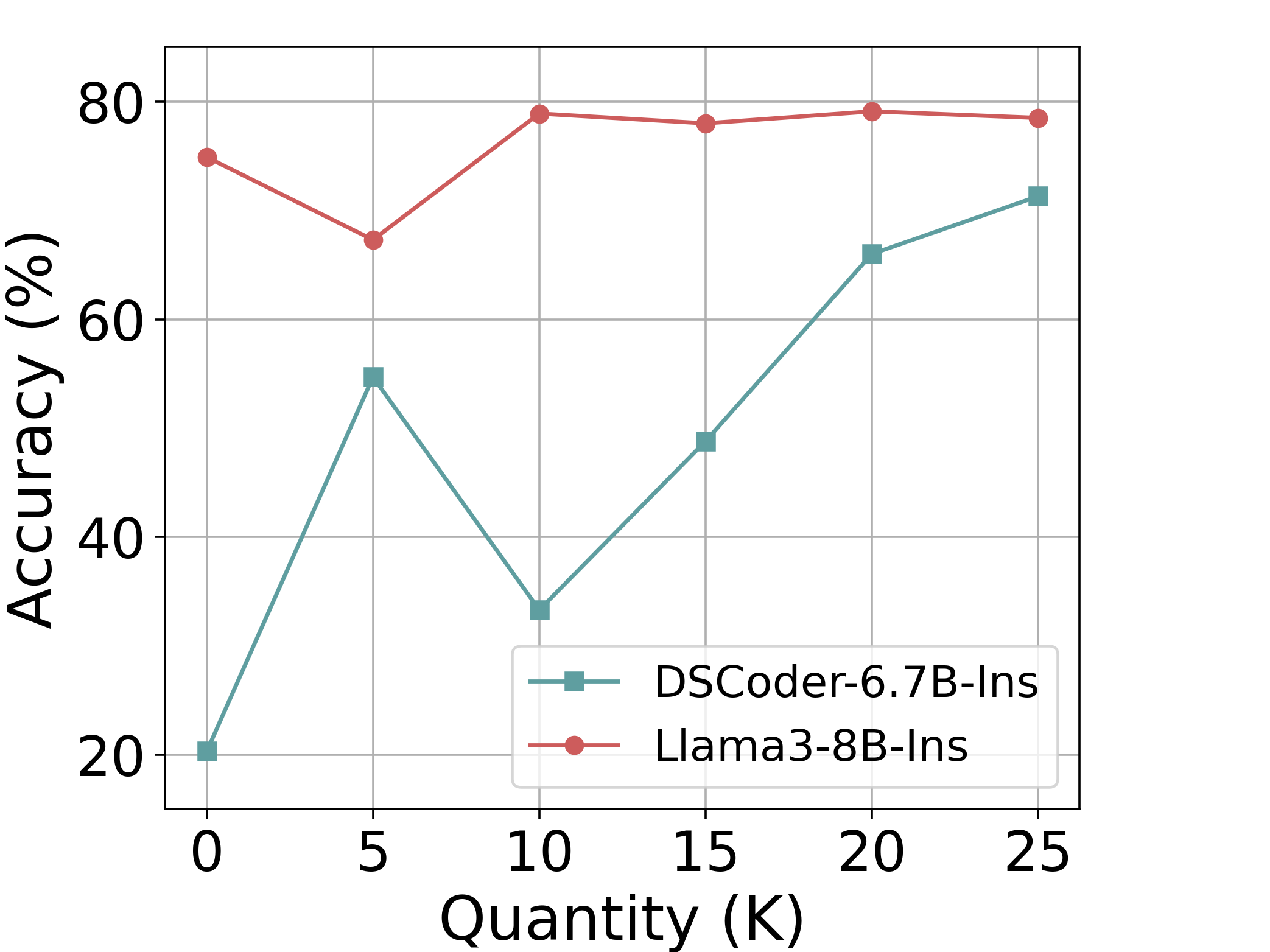}}
    \caption{Impact of the Amount of Training Data on Model Performance.}
    \label{fig:Analysis 2}
\end{figure}
\begin{table}
\small
\centering
\caption{Response Distributions of LLMs in BUG Identification Task. A.Syntax, B.Ref, C.Logic and D.Multi are four choices of the task.}
\label{tab:response distrubute}
\resizebox{\linewidth}{!}{
\begin{tabular}{l|cccc}
\hline
\multirow{2}{*}{\textbf{Model}} & \multicolumn{4}{c}{\textbf{Response Distributions (\%)}}          \\ \cline{2-5} 
 & \textbf{A.Syntax} & \textbf{B.Ref} & \textbf{C.Logic} & \textbf{D.Multi} \\ \hline
\multicolumn{5}{c}{\textit{Zero-Shot}}                                                                 \\ \hline
\multicolumn{1}{l|}{Llama3-8B-Ins}          & 0.0    & 0.8       & 71.0    & 28.2     \\
\multicolumn{1}{l|}{NeuDebugger-Llama3-8B}  & 11.4   & 3.9       & 8.6     & 76.1     \\
\multicolumn{1}{l|}{DeepSeek-Coder-V2}      & 28.0   & 7.2       & 36.8    & 28.0     \\ \hline
\end{tabular}}
\end{table}
\subsection{The Impact of Data Quantity}\label{app:data_quantity}
This subsection explores the impact of data quantity when finetuning LLMs using the NeuDebugger strategy. As shown in Figure~\ref{fig:Analysis 2}, we finetune the DSCoder-6.7B-Ins and Llama3-8B-Ins models using varying amounts of SFT data points. We then assess their performance on the \benchmark and visualize the results.

Compared to Llama3-8B-Ins, DSCoder-6.7B-Ins shows a significant performance increase when more SFT data are fed. This indicates that code-oriented LLMs are better at learning from debugging data, whereas a standard language model struggles to enhance its debugging capabilities without an essential understanding of code. Across all debugging tasks defined by \benchmark, DSCoder-6.7B-Ins exhibits significant improvements in BUG Localization, BUG Identification, and Code Recognition, while only showing slight improvements in the code repair task. This suggests that these debugging data do indeed contribute to the better code-repair ability of LLMs, though the task remains challenging to improve significantly.

\begin{figure*}[t] \centering
    \includegraphics[width=\textwidth]{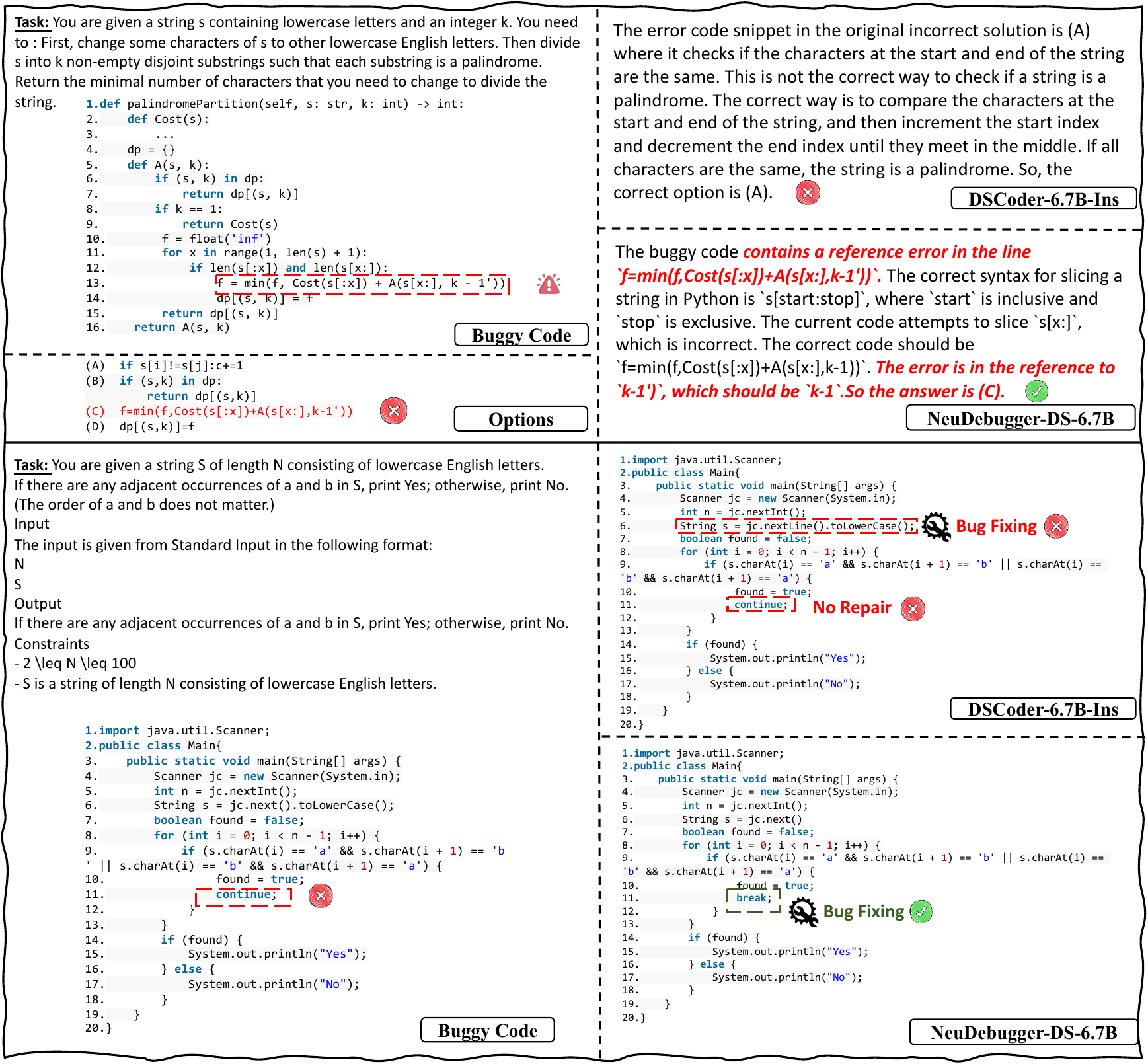}
    \caption{Case Studies. We provide two cases from BUG Localization task and Code Repair task to show the effectiveness of NeuDebugger.} \label{fig:case}
\end{figure*}
\subsection{Analysis of the Responses of LLMs in the Bug Identification Task}\label{sec:appendix:analysis}
In Table \ref{tab:response distrubute}, we show the response distributions of different models in the Bug Identification task.


The evaluation results show that LLama3-8B-Ins demonstrates a tendency to favor Logical Error and Multiple Errors. In contrast, DeepSeek-Coder-V2 performs in a more balanced manner compared to 7B-scale models, effectively avoiding random guessing behavior. In addition, NeuDebugger-LLama3-8B has a more even distribution of error types than the base model LLama3-8B-Ins, and is able to identify Syntax Error and Reference Error, which highlights NeuDebugger's ability to better identify error types rather than randomly guessing answers as Logical Error or Multiple Errors.

\subsection{Case Studies}
\label{sec:appendix:case_studies}
As shown in Figure \ref{fig:case}, we compare the performance of the DSCoder-6.7B-Ins model before and after training through two cases to demonstrate the effectiveness of NeuDebugger.

For the first case of the BUG Localization task, the code error is caused by the line \texttt{f = min(f, Cost(s[:x]) + A(s[x:], k-1'))}, which has an incorrect index \texttt{k-1'}. Thus, the correct answer is (C). DSCoder-6.7B-Ins considers the code fragment \texttt{if s[i]!=s[j]: c+=1} as erroneous, stating ``This is not the correct way to check if a string is a palindrome''. 
In contrast, NeuDebugger-DS-6.7B accurately analyzes
the reason for the bug ``contains an error in the line \texttt{f = min(f, Cost(s[:x]) + A(s[x:], k-1'))}, the error is in the reference to \texttt{k-1'}, which should be \texttt{k-1}'', demonstrating its effectiveness in BUG Localization.

In the second case of the Code Repair task, the error involves the misuse of \texttt{continue}, which leads to a Logical Error. The DSCoder-6.7B-Ins model fails to identify this error and instead suggests changing the line \texttt{String s = jc.next().toLowerCase()} to \texttt{String s = jc.nextLine().toLowerCase()}. This modification introduces a new error, as it does not handle the input correctly. NeuDebugger-DS-6.7B accurately recognizes that the problem lies in the use of \texttt{continue} and changes \texttt{continue} to \texttt{break}, successfully resolving the bug.

\begin{figure}[t] \centering
    \includegraphics[width=0.49\textwidth]{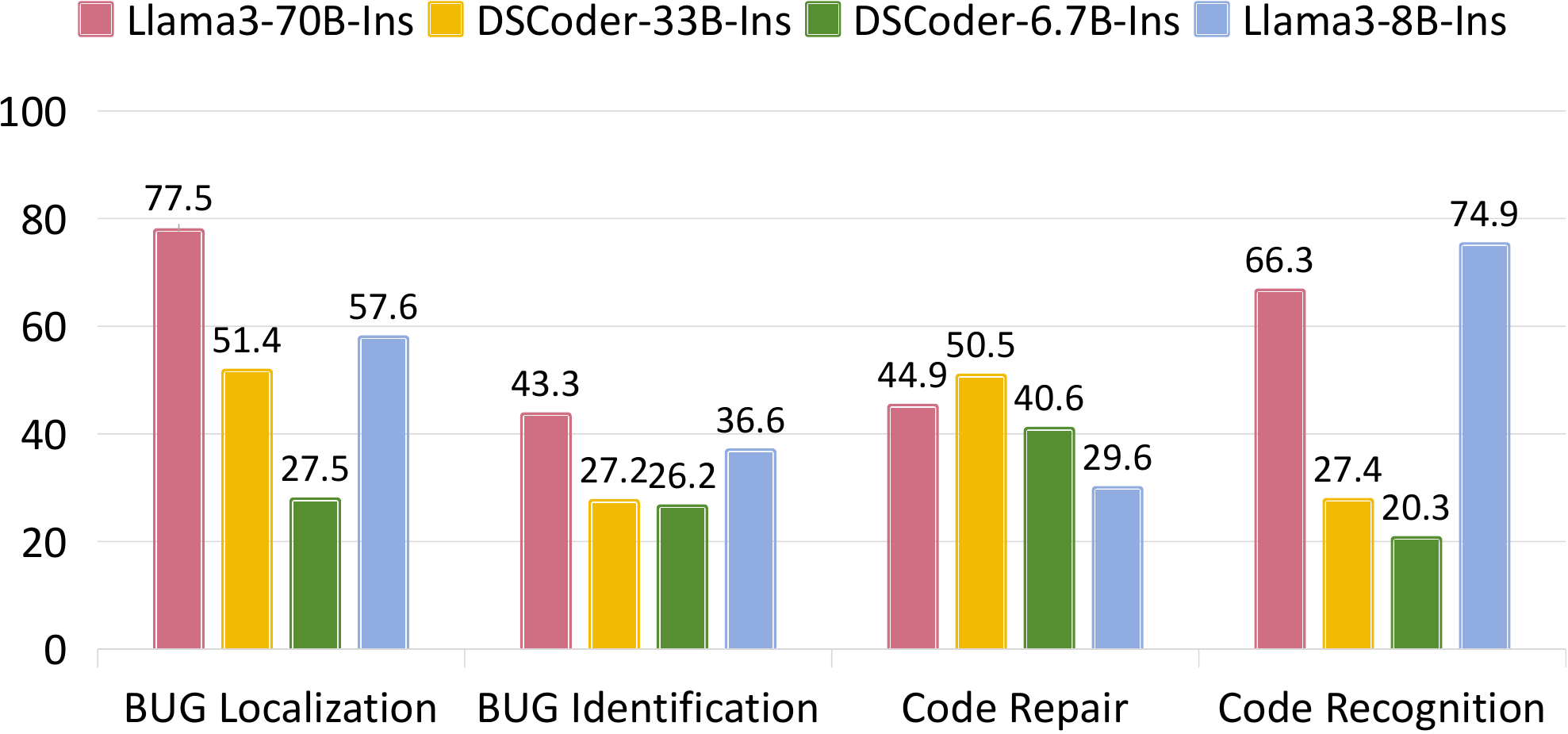}
    \caption{Debugging Performance of Different LLMs on \benchmark.} 
    \label{fig: model eval}
\end{figure}

\subsection{More Analyses of Tasks Defined in the \benchmark}
\label{sec:appendix:reason of four tasks}
This subsection provides an in-depth analysis of the tasks defined in \benchmark.

\textbf{Performance of Different LLMs.} As illustrated in Figure~\ref{fig: model eval}, we compare the performance of various LLMs in different tasks. The results reveal that their performance is inconsistent across tasks. For instance, while DSCoder-6.7B-Ins outperforms Llama3-8B-Ins in the Code Repair task, Llama3-8B-Ins achieves better results in Bug Localization, Bug Identification, and Code Recognition tasks. These findings illustrate the limitations of evaluating LLMs' code debugging capabilities based solely on a single Code Repair task, highlighting the importance of incorporating different tasks for evaluation.

\textbf{The Feasibility of Different Tasks.} The benchmark includes four task scenarios—Bug Localization, Bug Identification, Code Recognition, and Code Repair—that evaluate the debugging capabilities of LLMs. These tasks closely align with the skills demanded of developers in real-world scenarios. In the software development workflow, a developer typically identifies the buggy code, analyzes its root cause, implements a solution, and verifies the fix to ensure correctness. Thus \benchmark designs these tasks to mirror the key stages of real-world debugging.

The three tasks defined in \benchmark—BUG Localization, BUG Identification, and Code Recognition—are structured as single-choice question-answering problems, enhancing the accuracy of evaluation. This design is well-suited to the nature of these tasks. First, in real-world scenarios, code error localization involves pinpointing specific lines within faulty code, a process that naturally aligns with a single-choice problem. Second, BUG Identification requires classifying errors such as Syntax Error, Reference Error, Logical Error, and Multiple Errors, making it intuitive and efficient to adopt a single-choice format where the model selects the correct error type. Lastly, Code Recognition involves identifying the incorrect code fragment between two options, inherently fitting the single-choice question format.

\end{document}